# Performance Analysis of Synchronous Motor Drives under Concurrent Errors in Position and Current Sensing

**Prerit Pramod**, *Senior Member*, IEEE

Control Systems Engineering, MicroVision, Inc.
*Email*: preritpramod89@gmail.com; preritp@umich.edu; prerit_pramod@microvsion.com

**Abstract**: Field oriented control of permanent magnet synchronous motor drives involves the closed-loop regulation of currents in the synchronous reference frame. The current feedback is directly affected by errors in both position and stationary frame current measurements. This paper presents the exact analytical expression for estimated synchronous frame currents under simultaneous errors in both sensors along with a detailed analysis of the incorrect estimation on the closed-loop current control performance.

## Introduction

Closed-loop current control of permanent magnet synchronous motors (PMSMs) is performed by utilizing a current regulator that minimizes the error between the commanded and measured (estimated) synchronous ($dq$) frame currents, which are in-turn computed by applying the Park transform involving an estimate of the electrical position on the stationary ($abc$) frame currents [1]–[3]. Thus, erroneous position or current measurements significantly degrade the closed-loop behavior current tracking behavior and also cause torque ripple. Since the transformations are non-linear, models that account for either only position sensing errors [4]–[7], or only current sensing errors [8]–[15], prevalent in existing literature, do not explain the behavior under simultaneous errors because they exhibit modulation of distinct harmonic frequencies caused by the individual error sources.

This paper provides a summary of generalized mathematical models that capture the effects of both position and current sensing errors simultaneously on the estimation of synchronous frame currents for PMSM drives [16]. The model is validated with experimental results on a practical PMSM drive and is then used to study the behavior of systems in detail. The universality of the modeling approach expands its applicability to drive systems employing other electric machine topologies including, among others, induction, (other) synchronous and switched reluctance motors [17]–[21]. The model has widespread utility as it enables the development of control algorithms for detection, isolation and mitigation of simultaneous errors in multiple transducers, i.e., position and current.

## Overview of Permanent Magnet Synchronous Motor Drives

The current control system of a typical PMSM drive system is shown in Fig. 1. The current regulator applies voltage commands based on the synchronous frame current commands and the corresponding current estimates in order to minimize the current tracking error. The synchronous frame voltage commands are converted to duty cycles by the inverter commutation block using space vector pulse width modulation (SVPWM) and applied to the motor via the gate driver and inverter [22]–[25]. Under ideal conditions, i.e., no sensing errors, the output current matches the estimated currents and thus the commands.

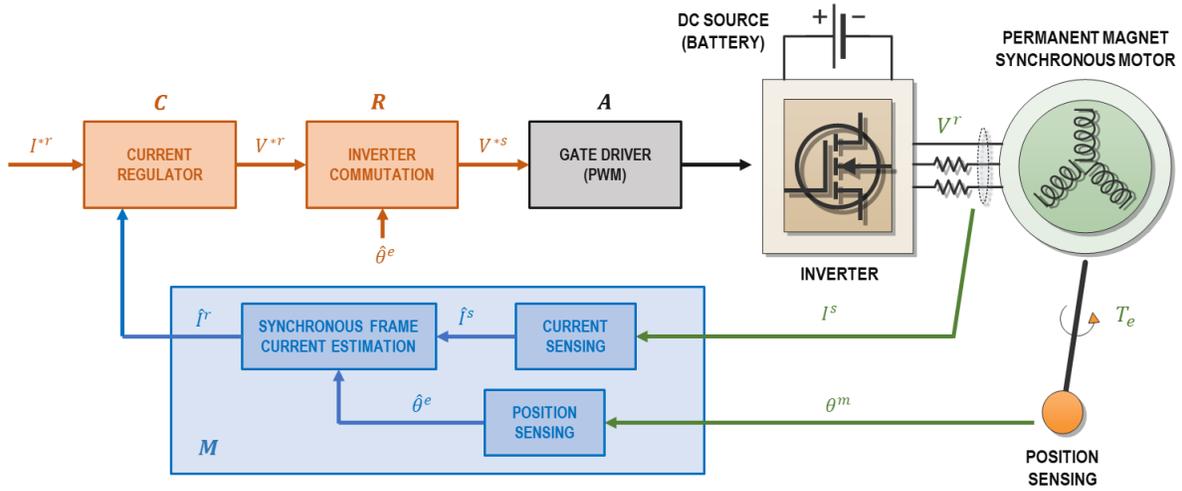

**Fig. 1.** Current controlled PMSM drive system.

The analytical model of a PMSM in the synchronous frame with an ideal inverter and negligible transport delay due to discrete PWM is expressed in (1) [26]–[28].

$$V^r = P^{-1}I^r + E^r$$
$$\begin{bmatrix} V_d \\ V_q \end{bmatrix} = \begin{bmatrix} L_d s + R & \omega_e L_q \\ -\omega_e L_d & L_q s + R \end{bmatrix} \begin{bmatrix} I_d \\ I_q \end{bmatrix} + \begin{bmatrix} 0 \\ \omega_e \lambda_m \end{bmatrix} \quad (1)$$

where $I^r$ and $V^r$ are the (actual) synchronous frame voltages and currents respectively, $E^r$ is the back-EMF term, $P$ is the machine transfer matrix, $L_d$ and $L_q$ represent the direct and quadrature axis inductances respectively, $R$ represents the combined motor and inverter resistance, $\lambda_m$ is the permanent magnet flux linkage and $\omega_e$ is the electrical velocity or synchronous frequency of the machine.

## Synchronous Frame Current Estimation

Since synchronous frame current estimation involves the conversion of measured stationary frame currents $\hat{I}^s$ through the Park transformation $H$ which in-turn utilizes estimated electrical position $\hat{\theta}^e$, errors in current and position estimation results in erroneous estimated synchronous frame currents $\hat{I}^r$ as expressed in (2).

$$\hat{I}^r = H(\hat{\theta}^e)\hat{I}^s = H(\hat{\theta}^e)\left((J + \overline{K}_g)I^s + \overline{K}_o\right) \tag{2}$$

where $\hat{I}^r$ and $\hat{I}^s$ are the estimated synchronous and stationary frame currents respectively, $I^s$ is the actual stationary frame current, $J$ is the identity matrix, $\hat{\theta}^e$ is the electrical position estimate, while $\overline{K}_g$ and $\overline{K}_o$ are the current gain and offset error matrices as given in (3).

$$\overline{K}_g = \begin{bmatrix} \Delta K_a & 0 & 0 \\ 0 & \Delta K_b & 0 \\ 0 & 0 & \Delta K_c \end{bmatrix} \qquad \overline{K}_o = \begin{bmatrix} \Delta I_a & 0 & 0 \\ 0 & \Delta I_b & 0 \\ 0 & 0 & \Delta I_c \end{bmatrix} \tag{3}$$

where $\Delta K_x$ and $\Delta I_x$ represent gain and offset errors respectively in phase $x$. It should be readily inferred that the synchronous frame current vector consists of the $d$, $q$ and 0 sequence currents. The electrical position estimate $\hat{\theta}^e$ is related to the true position $\theta^e$ as (4).

$$\hat{\theta}^e = \theta^e + \Delta\theta^e \qquad \Delta\theta^e = \delta\theta_0^e + p\sum_r \delta\theta_r^m \qquad \delta\theta_r^m = \theta_{ra}\sin(r\theta^m + \phi_r) \tag{4}$$

where $p$ is the number of magnetic pole pairs, $\Delta\theta^e$ is the total position error consisting of static (constant) errors $\delta\theta_0^e$ and harmonics $\delta\theta_r^m$. The mechanical order $r$ of the harmonic depends on the type of error in position estimation.

Since the two estimation error relationships are non-linear, the synchronous frame consists of errors due to the individual sources as well as a combination consequently. The final relationship between the estimated and true synchronous frame currents obtained by combining (1), (2) and (4) is given in (5).

$$\hat{I}^r = M_\theta I^r + M_{ig}I^r + M_{io} = M_\theta I^r + \left(K_{igc}M_\theta + K_{igp}M_{igp}\right)I^r + K_{iop}M_{iop} \tag{5}$$

where $M_\theta$ is the position estimation error matrix while $M_{ig}$ and $M_{io}$ are the current measurement gain and offset error matrices respectively. The gain error matrix is further expressed as a sum of the position error $M_\theta$ matrix and an additional matrix $M_{igp}$ specific to gain errors, scaled by constants $K_{igc}$ and $K_{igp}$ respectively. These matrices and constants are directly presented as (6) with the zero-sequence current omitted (for compactness).

$$M_\theta = \begin{bmatrix} \cos\Delta\theta^e & \sin\Delta\theta^e \\ -\sin\Delta\theta^e & \cos\Delta\theta^e \end{bmatrix} \qquad M_{iop} = \begin{bmatrix} \cos(\theta^e + \phi_{iop}) & \sin(\theta^e + \phi_{iop}) \\ -\sin(\theta^e + \phi_{iop}) & \cos(\theta^e + \phi_{iop}) \end{bmatrix}$$

$$M_{igp} = \begin{bmatrix} \cos(\theta^e + \hat{\theta}^e + \phi_{igp}) & \sin(\theta^e + \hat{\theta}^e + \phi_{igp}) \\ -\sin(\theta^e + \hat{\theta}^e + \phi_{igp}) & \cos(\theta^e + \hat{\theta}^e + \phi_{igp}) \end{bmatrix} \tag{5}$$

$$K_{igp} = \sqrt{\sum_x \Delta K_x^2 - \sum_{x,y;x\neq y} \Delta K_x \Delta K_y} \qquad K_{igc} = \frac{1}{3}\sum_x \Delta K_x \qquad K_{iop} = \frac{2}{3}\sqrt{\sum_x \Delta I_x^2 - \sum_{x,y;x\neq y} \Delta I_x \Delta I_y}$$

The error matrices shown in (5) clearly illustrate the non-linear modulation of harmonics due to simultaneous errors in both position and current estimates. The position error matrix $M_\theta$ may be static or consist of (mechanical) order-tracked pulsating terms depending on the type of position error. The error term $K_{igc}M_\theta$ incorporates the effect of simultaneous current sensor gain and position error, i.e., the term may be static or dynamic due to the type of position error but its magnitude is dependent on the amplitude of gain errors. Stationary frame current sensor offset errors translate to first electrical order harmonics in the synchronous frame but may further exhibited frequency modulation based on the type of position estimation error due to their dependency on estimated rather than true position.

### Closed-loop Current Control

A simple feedback regulator used for synchronous frame current command tracking consisting of dual PI controllers and feedforward disturbance (back-EMF) compensation along with the synchronous frame current estimation including errors is shown in Fig. 2.

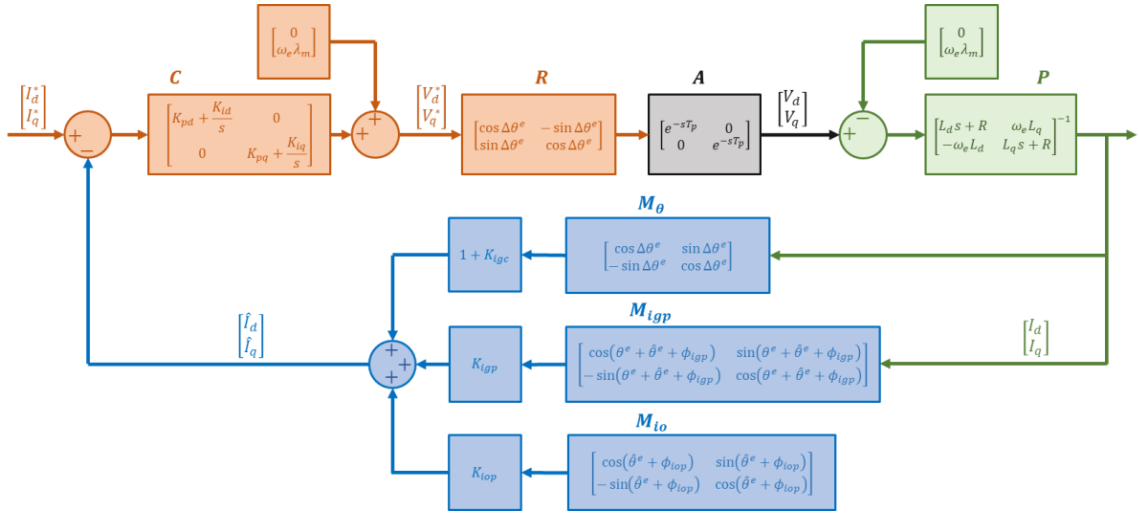

**Fig. 2.** Synchronous frame current regulator.

Although the exact closed-loop dynamics depend on the structure and tuning of the current controller, a simplified expression for the actual currents may be obtained if ideal command tracking is assumed as with high bandwidth tuning as shown in (8) [29]–[35].

$$\hat{I}^r \approx I^* \quad \rightarrow \quad I^r \approx \left(M_\theta + M_{ig}\right)^{-1}(I^* - M_{io}) \tag{8}$$

The control signal, i.e., voltage command output from the regulator may be obtained using the system dynamics. The detailed analysis of the effects of erroneous synchronous current estimation on closed-loop current control is excluded from the digest for brevity.

## Model Validation

The analytical model presented is validated experimentally on a 9-slot, 6-pole non-salient pole PMSM with $R = 0.0106\ \Omega$, $L_d = L_q = 59.45\ \mu\text{H}$ and $\lambda_m = 0.0077$ Wb. The experimental setup (not shown) includes an induction motor servo driving the test motor at a constant speed of 100 RPM while open-loop current control operation is enabled in order to observe the estimated currents as shown in Fig. 3 without the influence of the regulator with two current commands $I_q^* = 0, 21.5$ A (with $I_d^* = 0$ A) and different error combinations are injected.

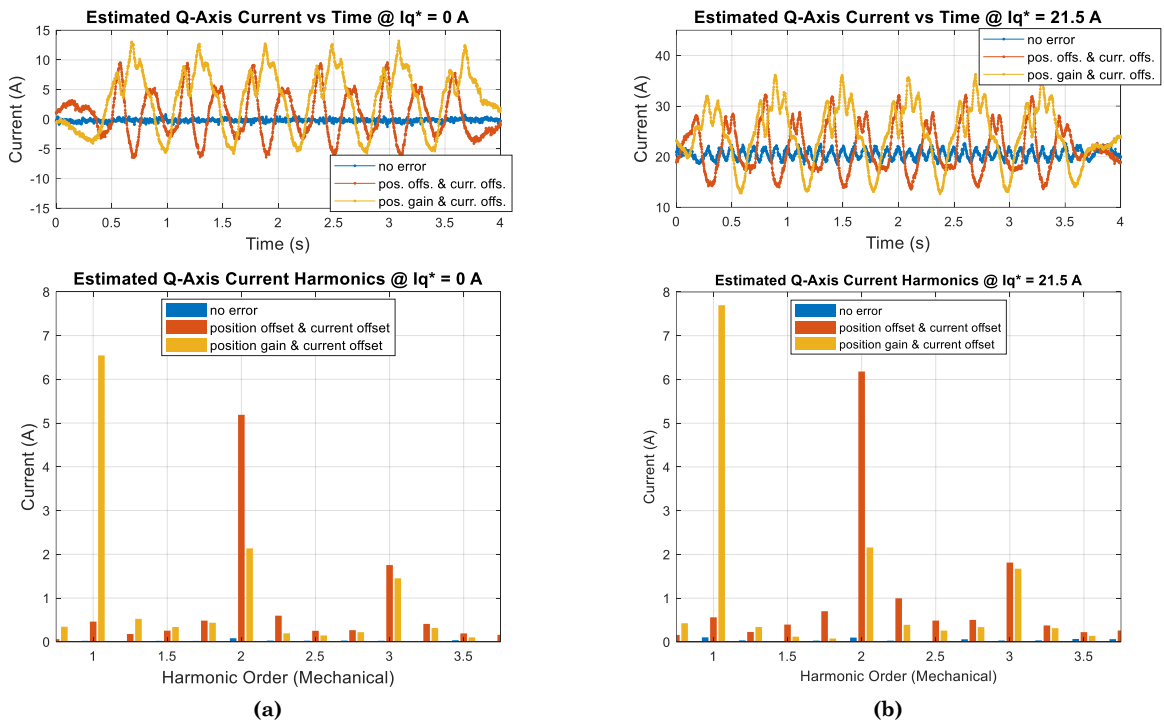

**Fig. 3.** Experimental current estimates vs time and harmonics at current levels (a) 0 A and (b) 21.5 A at constant speed of 100 RPM with simultaneous position and current sensing errors under open-loop operation.

The experimental results clearly illustrate the spectral spreading around the base orders as predicted by the analytical results, thus validating the model presented.

## Conclusions

A novel generalized mathematical model describing the effects of simultaneous position and current sensing errors on synchronous frame current estimation and closed-loop current control of PMSM drives, validated with experimental results, is presented in this paper.